\documentclass{ceurart}
\usepackage{soul}
\usepackage{amssymb}
\usepackage{pifont}

\usepackage{multicol}
\usepackage{graphicx}%
\usepackage{multirow}%
\usepackage{amsthm}%
\usepackage{mathrsfs}%
\usepackage{xcolor}%
\usepackage{textcomp}%
\usepackage{manyfoot}%
\usepackage{booktabs}%
\usepackage{algorithm}%
\usepackage{algorithmicx}%
\usepackage{algpseudocode}%
\usepackage{listings}%
\usepackage{listings}
\usepackage[acronym]{glossaries}
\usepackage[inline]{enumitem}
\usepackage{subcaption}

\begin{document}

\copyrightyear{2024}
\copyrightclause{Copyright for this paper by its authors. Use permitted under Creative Commons License Attribution 4.0 International (CC BY 4.0).}
\conference{6th  International Conference in Recent Trends and Applications in Computer Sciences and Information Technologies}

\title{Elevating Semantic Exploration: A Novel Approach Utilizing Distributed Repositories}

\author[1]{Valerio Bellandi}[%
  orcid=0000-0003-4473-6258,
  email=valerio.bellandi@unimi.it,
]
\fnmark[1]

\address[1]{Università Degli Studi di Milano, Department of Computer Science, Via Celoria 18, Milano, Italy}

\cortext[1]{Corresponding author.}
\fntext[1]{These authors contributed equally.}

\begin{abstract}
Centralized and distributed systems are two main approaches to organizing ICT infrastructure, each with its pros and cons. Centralized systems concentrate resources in one location, making management easier but creating single points of failure. Distributed systems, on the other hand, spread resources across multiple nodes, offering better scalability and fault tolerance, but requiring more complex management.

The choice between them depends on factors like application needs, scalability, and data sensitivity. Centralized systems suit applications with limited scalability and centralized control, while distributed systems excel in large-scale environments requiring high availability and performance.

This paper explores a distributed document repository system developed for the Italian Ministry of Justice, using edge repositories to analyze textual data and metadata, enhancing semantic exploration capabilities.

\end{abstract}


\maketitle
\section{Introduction}\label{secintro}
In the digital age, data has become the foundation of innovation, driving progress in various industries, from finance to healthcare and beyond. However, as the volume and complexity of data grow, so do concerns about security and privacy. Furthermore, data plays a pivotal role in modern information and communication technology (ICT) infrastructure, shaping decision-making, enhancing efficiency, and fostering technological advancements. The role of data in ICT systems is indispensable, as it forms the core framework upon which contemporary technologies are built and refined.  

ICT infrastructures can be structured using centralized or distributed models, each offering distinct benefits and challenges. In a centralized system, all computational resources and data processing occur within a single location or data center. This setup simplifies maintenance and management while providing direct control over operations and data governance. However, it comes with certain drawbacks, such as potential bottlenecks, scalability limitations, and vulnerability to single points of failure. Additionally, accessing data from geographically distant locations can introduce latency issues.  

On the other hand, distributed systems disperse processing power and resources across multiple nodes or locations, enhancing scalability, resilience, and fault tolerance. This decentralized model supports efficient data processing, reduces latency, and mitigates failures through redundancy. However, distributed architectures introduce added complexity, requiring sophisticated coordination mechanisms to maintain data consistency and ensure system reliability.  

Choosing between centralized and distributed architectures depends on multiple factors, including application requirements, scalability demands, geographical distribution of users, and data sensitivity. While centralized solutions may be preferable for applications with modest scalability needs and strict data governance, distributed systems are better suited for large-scale deployments requiring high availability, robustness, and performance efficiency.  

This paper explores a real-world application involving a distributed document repository and metadata management system. Our proposed solution comprises a network of edge repositories that analyze textual documents and metadata to identify key entities. The primary objective is to introduce advanced semantic exploration functionalities for the Italian Ministry of Justice.  

\section{State of the Art}\label{secstate}

Several systems have been proposed for legal document management. For example, \cite{PAUZI2023111616} reviews software architectures for NLP in legal documents, identifying approaches like pipeline, service-oriented, and microservices architectures, with a focus on pipeline systems. While service-oriented and microservices architectures offer advantages, the study doesn’t propose a generic infrastructure for managing legal entities. The proposal in \cite{amato2008} presents a system combining NLP and ontologies for managing paper documents, converting them into RDF statements for indexing, retrieval, and preservation.
This system shares similarities with our district design but doesn’t emphasize entities as our system does. The work in \cite{humphreys2020} describes a knowledge management system that semi-automates the extraction of norms for legal ontologies using NLP modules and domain-specific rules. In contrast, \cite{BELLANDI2024105904} introduces an entity-centric architecture for managing court judgments and legal documents, similar to the district architecture in section. Our work extends this by introducing the concept of hierarchy. Other systems, such as \cite{QIN2024105930} and \cite{Ridwandono_Afandi_Wahyuni_Simaremare_Sinaga_2023}, focus on storing and organizing legal documents, but lack analytical capabilities. From an architectural perspective, distributed systems have been explored from various angles, including design, development, deployment, and non-functional behavior evaluation. Recent research emphasizes big data architectures, particularly in the edge-cloud continuum, focusing on performance and scalability (e.g., \cite{10.1177/1094342019877383,9193894}). Additionally, the impact of distributed systems on safety, security, and privacy has been considered, especially in terms of trustworthiness, governance, risk, and compliance. Assurance techniques \cite{AADV.CSUR2015} are now used to verify non-functional properties (e.g., availability, confidentiality, privacy) in distributed systems, with certification recognized as the primary method for ensuring these qualities \cite{AB.SSE2023}.

On the other hand, many studies refer to specific aspects of the legal document analysis and NLP applications. A good overview can be found in~\cite{zhong2020does}, which emphasizes the role of Named Entity Recognition (NER) techniques and Relation Extraction (RE). Usage of ontologies and of widely used NLP models like BERT in the legal domain has been reported (e.g. \cite{chalkidis-etal-2020-legal}). NLP methods have been applied to support legal information extraction and retrieval (see e.g. \cite{CASTANO2022101842}); contributions to the Competition on Legal Information Extraction/Entailment (COLIEE), organized since 2017~\cite{cooliedescropt}, describe several studies in this area. 


\section{The Data Model}\label{secdmodel}

The system consists of multiple local instances, all sharing the same architecture outlined in Section \ref{secarch}, along with a top-level instance. These local instances can be arranged hierarchically in a multi-level tree structure. However, in many cases, a simpler two-level configuration—comprising district-level local instances and a top-level instance—is sufficient.

\begin{figure}[h]
\centering
\includegraphics[width=0.7\textwidth]{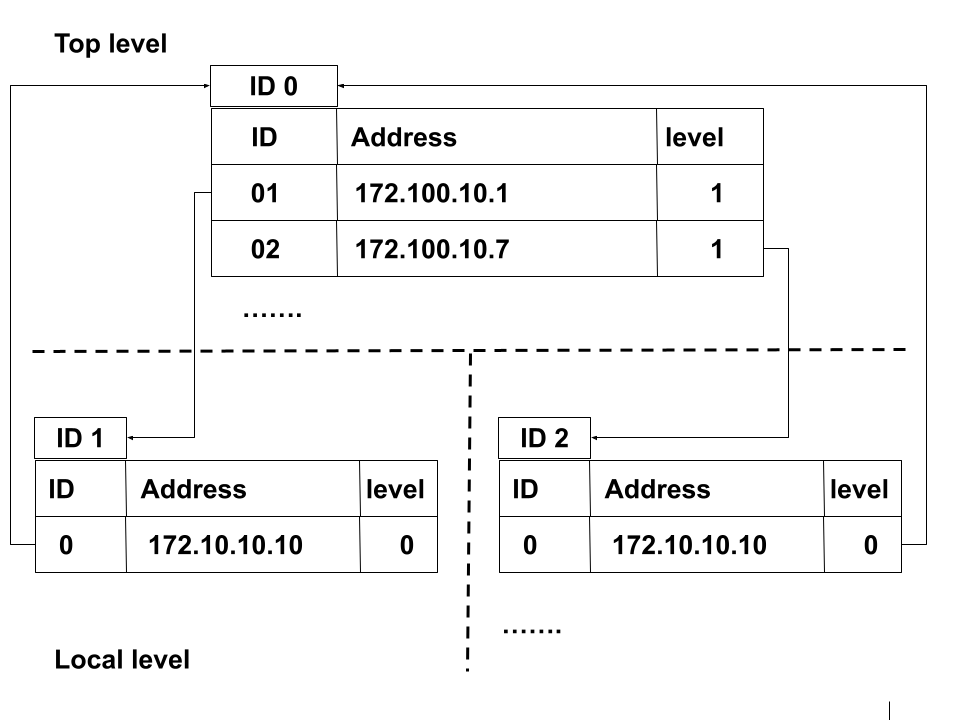}
\caption{Illustration of the instances hierarchy}\label{fighier}
\end{figure}

At each level, the system maintains a table containing the addresses and identifiers of lower-level instances, along with a reference to its parent instance. The table entries follow the format $\langle IID, A, L \rangle$, where $IID$ represents the instance identifier, $A$ denotes its address, and $L$ indicates its hierarchical level relative to the current instance. For a visual representation, refer to Figure \ref{fighier}.

Unique instance identifiers are generated by the top-level instance upon creation and remain distinct throughout the entire system.

As detailed in the upcoming sections, some data is centrally managed to ensure overall system consistency, while other information is distributed and stored locally. In the latter scenario, multiple instances may hold duplicate copies of the same data, each tagged with the identifier of the instance that maintains it.

\subsection{Documents: Text and Metadata}\label{subsecdmodel2}

The dataset comprises natural language textual documents, specifically court decisions, along with their associated metadata. Each document, denoted as \( d \), is represented as a triple \( d = \langle IID, M, T \rangle \), where \( IID \) is the identifier of the instance storing the document, \( M \) is the set of metadata, and \( T \) contains the document’s textual content. A metadata item \( m \in M \) is expressed as \( m = (mn_i, mv_i) \), where \( mn_i \) represents the metadata name and \( mv_i \) its corresponding value. Examples of metadata include the case number, the year of the decision, the presiding judge’s name, and similar attributes.  

The textual content is divided into sections. Formally, a section \( s \in T \) is defined as \( s = (sn_i, sv_i, sc_i) \), where \( sn_i \) is the section name, \( sv_i \) contains the full text of that section, and \( sc_i \) represents a collection of \emph{chunks} derived from segmenting the section’s content. Sections typically include elements such as the preamble (identifying the involved parties and the court), the case summary, and the final ruling.  

Each section can be further divided into chunks, which serve as a tokenized representation of its text. These chunks may be predefined blocks of characters or correspond to paragraph divisions within the section. Reconstructing a section by combining its chunks \( sc \) results in the original section content \( sv \).  

Documents can be duplicated and stored across multiple instances. In such cases, the textual content remains unchanged across copies, while the metadata, section structure, and tokenization may vary between instances. When a document is replicated, its metadata is transferred along with it, and the instance identifier of the new copy is updated accordingly.

\subsection{Annotations}\label{subsecdmodel3}
In general terms, an {\bf annotation} is the association of a document segment with a tag. An annotation $a$ is defined as $a = \langle IID, d, t, [e], (start;end) \rangle$, where $IID$ is the identifier of the instance where it is stored, $d$ is the reference to a document, $t$ is the tag and $e$ is the optional reference to an entity (see next section), and $(start;end)$ is the tagged segment delimited by positions $start$ and $end$ within the document $d$.
In classical Named Entity Recognition tasks, annotations are used to tag a document with entity types, for instance persons, organizations and so on.

We extend the use of annotations to refer text portions to real entities, for instance a specific individual, not just a person.

It is intended that referred entities are stored in the same instance as their annotations and that, if an annotation is copied to another instance, all the referred entities are also copied.

\begin{figure}[h]
\centering
\includegraphics[width=0.50\textwidth]{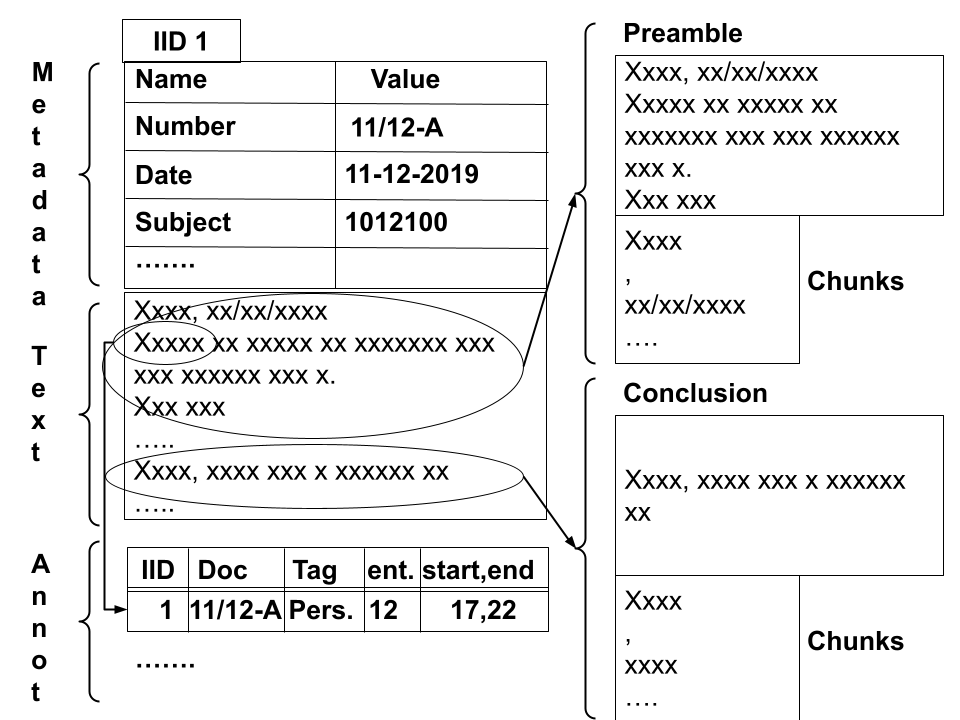}
\caption{Illustration of a document with sections and annotations}\label{figtext}
\end{figure}

The situation is illustrated in fig. \ref{figtext}, where a document stored at \textit{IID 1} is represented. On the left some metadata, the text and an annotation can be found. It is assumed that the document can be identified by its number; the annotation refers to a person (tag \textit{Pers.}), having ID 12 in the Entity Register (see section \ref{subsecdmodel4}). On the right, two sections are represented: the \textit{Preamble} and \textit{Conclusion}. For both, in this example, chunks consist of the set of words.

\subsection{Entities}\label{subsecdmodel4}
An entity $e$ is an object found in a document. It is defined as a triple $e = \langle IID, E, F \rangle$, where $IID$ is the identifier of the instance where it is stored, $E$ is an entity type, and $F$ is a set of attributes with related values that qualifies the entity $e$ according to the type $E$. 

An {\em entity type} $E$ describes an object template and it is defined as $E = (en, ef, ek)$, where $en$ is the entity name, $ef$ is a set of features that characterize the entity type, and $ek$ is a set of keys, namely combinations of features within $ef$ whose values uniquely identify the entity.
For instance, the entity type person can be identified by the set \textit{name, surname, date of birth, place of birth} and, in Italy, by the \textit{fiscal code} assigned by the State.

Entity type definitions constitute a common resource, therefore they are managed at the top level instance and may be queried by any local instance as needed.

We note that entities may be of abstract nature too, for instance laws and articles, concepts and so on.

The collection of entities at any levels is the \textit{Entity Register} (EReg) and the definition of entity types is its \textit{Metamodel}. The goal of the whole system is to make information about entities available at all districts, according to user access rights. A major issue with this goal is the fact that entities, in addition to having different ids in the local ERegs, may have been identified using different attributes.
For instance, a person may have been identified by \textit{Name, Surname, Birth Date, Birth Place} in an EReg and by \textit{Name, Surname, Mother Name, Mother Surname, Father Name, Father Surname} in another.

For this reason, we use a top level EReg, and implement the structure of both the top level and local ERegs in order to store i) all the entities' attributes ii) the entities' relationships iii) ids of entities in local ERegs. In other terms, the top level EReg collects available information about entities, helping users to merge or disambiguate entities.

About \textit{attributes}, the EReg metamodel is implemented as follows: i) it stores also attributes that do not make part of any identifiers ii) a value type in the attribute definition, that can be used to check the attribute validity. It also distinguishes cases when a single value or multiple cases are allowed. Formally, $A= \langle an, at \rangle$, where $an$ is the attribute name and $at = text | integer | float | date | ... | list$ is the valid value type. In particular, only when type is list an entity can locally have different values. For instance the attribute \textit{Eyes Color} may have just one value for a person: if two different values are found for an entity, either it must be split or there is an error in the data. On the contrary, the attribute \textit{qualification} may have several values for the same individual. About \textit{relationships}, the EReg metamodel stores: i) the involved entity types and the direction if any ii) the \textit{cardinality}, iii) optionally the relationship \textit{validity period} flag. The cardinality specifies whether only a fixed maximum number $M$ relationships of this type may exist between two entities, or any numbers. Formally, $R = \langle rn, set, tet, bid, cst, cts, vp \rangle$ where $rn$ is the relationship type name, $set$ is the type of the source entity, $tet$ is the type of the target entity, $bid$ is a flag to specify the characteristics of the relationship with regards to bidirectionality, that is if the relationship is bidirectional or mono directional and in this case if a relationship in the opposite direction may exist or is contradictory. $cst$ is the cardinality of targets, that is it specify if one source can be linked to only 1 target, to a number up to some $N$ or to any number; $cts$ is the analogous for sources. Finally, $vp$ is a flag specifying if the relationship may last after a while; .in such case the relationship instances have a start and an end date.
For instance the relationship \textit{Father Of} has \textit{person} as source and target entity type, is not bidirectional and contradictory with the opposite relationship; $cts=1$ (only one father is allowed), but $cst > 1$, and $vp = False$. 
\textit{Grandmother Of} has $cts=2$ and \textit{Friend Of} is bidirectional and has both $cts$ and $cst$ greater than 1. An example of a mono directional relationship that is not contradictory with its opposite is \textit{In Love With}.
A typical relationship having  cardinality 1 and a validity period is\textit{Married With}, as marriage may be interrupted by divorce or death of the mate.

Some relationship types are mutually exclusive, for instance a person cannot be at the same time \textit{Father Of}, \textit{Mather Of} and \textit{Granfather Of} somebody. Accordingly, the EReg stores relationships between the nodes representing contradictory relationships between the entities: $\langle R1, R2, CONTR \rangle$ where $R1$ and $R2$ are relationship types and $CONTR$ is the label of the relationship.

Finally, the EReg metamodel may be implemented with supplementary rules to i) deduce some attributes or relationships by others ii) specifying constraints. As an example of the first type, we consider the Italian personal tax code \textit{codice fiscale}, that may be used to deduce birth date and place. For the second type, we consider constraints like \textit{PhD date must be greater than birth date plus N years}.

We assume for the sake of the simplicity that entities, attributes, identifiers and relationships have the same names in all the ERegs.

\subsection{Data Access Constraints}
\label{ssecdac}
This section delves into the intricacies of data access, underscoring the diverse privacy concerns inherent in the types of entities outlined earlier. Notably, while personal information demands a stringent level of privacy protection, legal articles carry no such imperative. Consequently, access to queries pertaining to individuals should be restricted to select users, whereas access to information regarding legal articles should be available to all. However, documents retrieved as a result of querying an article often contain references to individuals, necessitating either their concealment from general users or anonymization. Permissions are managed through the following data structures.

The \textit{Entity Type Privacy} table has elements $\langle E, pl \rangle$ with $E$ entity type and $pl$ privacy level, from $0$ (public) to $N$ (highly private). This is a global table, stored in the top level instance.

The \textit{User - Document - Ownership} table, with entries $\langle IID, U, D, O \rangle$, where $IID$  is the identifier of the instance where the document is stored and the user can login, $U$ is a user, $D$ a document and $O$ an ownership level, for instance owner, editor, reader, and so on. A specific level is \textit{generic}, meaning that the user can see only sections of the document that do not contain entities, number of contained entities and other general information. Depending on the organization complexity, an alternative format is $\langle IID, Ug, Dg, O \rangle$, where $Ug$ and $Dg$ are respectively users and documents groups, and it is supposed that other tables relate each user and document to their groups.

The \textit{Privacy - Permission} table, has an entry for each combination of the values of tables above: $\langle O, pl, P \rangle$, where $O$ is an ownership level, $pl$ a privacy level and $P$ a permission level. For instance, it can be specified that document owners have full permission on any mentioned entities, but readers can see only entities up to level $M$, while entities of higher level must be anonymized. Noteworthy examples of $P$ are: full control, read only, read anonymized, without mentions (that is the user can see the entity but not its mentions in documents), count only (that is the user can only see how many entities/mentions of some types are in some documents, without any details).

\subsection{Top Level Functions}
\label{sectopl}
\begin{figure}[h]
\centering
\includegraphics[width=0.6\textwidth]{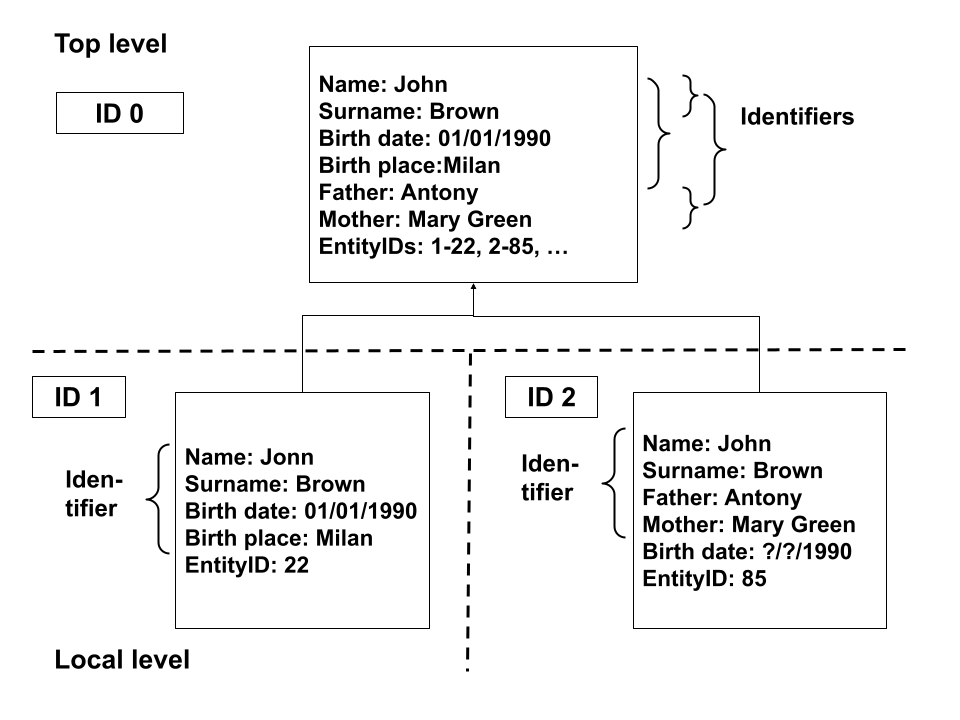}
\caption{Illustration of an entity entry at local and top level}\label{figtopent}
\end{figure}

With the data structures described in previous section, we can build at the top level enlarged versions of the entities including all the available information in the whole district network, while at the district level entities with the locally available information are maintained. The situation is illustrated in fig. \ref{figtopent}: in the local Entity Register at ID 1, a person is uniquely identified by name, surname, date and place of birth, and assigned ID 22. In the Entity Register at ID 2, a person exists with same name, surname and year of birth, and is assigned ID 85; such person is locally uniquely identified by name, surname, father and mother. Synchronization functions, possibly with the user help to resolve ambiguous cases, recognize that the person is actually the same and store at top level the complete set of data, including identifiers definition and IDs assigned at both local levels.

The enlarged entity management (both at the local and top level) may be performed either as soon as an entity is recognized in a document or at fixed times (e.g. once a day). At the top level, the entity synchronization might be even postponed at the time when queries are submitted.
We will first describe the main cases that may occur, in the hypothesis that an entity is managed as soon as it is recognized in a document, then we will discuss the other options.

Suppose that at \textit{District1} a document $D_1$ is analysed by a service and entity $E_1$ is found, with attributes $a_1,...,a_n$ that constitute an identifier $i_1$ other attributes $a_{n+1},...,a_N$ and relationships $r_1, ..., r_n$ to existing entities. The service will query the local EReg with $i_1$ to know if the entity exists. If it does not, an entry will be created for it, with all its attributes. The local EReg will also query the top level EReg, supposing that the entity is not found, a new entry will be created at the top level too; the name of the local entity and the assigned id is stored. All relationships with other entities are stored, both at local and top level; conflicts, if any, are submitted to the user; for instance a new entity could be quoted as father of somebody who already has a father.

When a second document $D_2$ is analysed at the same district and an entity with the same $i_1$ is found, an attribute set $a_1,...,a_n,$ $A_1,...,A_M$ and relationships $r_1, ..., r_n$ $R_1, ..., R_N$, the new attributes and relationships are compared to the other stored for the entity in the local EReg. If they are compatible, the attribute and relationship sets are enlarged; if not the user is requested to solve the conflict, e.g. fixing some attributes and relationships or splitting the entities. The top level EReg will be updated to add the new attributes or store the changes.

It may happen, however, that the new entity found lacks some attributes to compose any complete identifiers; for instance a person is found with only name, surname and secondary attributes. In this case, the service will query the local EReg with the partial identifiers and get a list of compatible entities with their attributes and relationships. Attributes and relationships compatibility is used to guess entities that might coincide with the one found; the user is prompted to choose an existing entity or to create a new one. In the first case new attributes, if any, are added to the old entity.
The data is then sent to the top level EReg, and compared to the larger available set of compatible entities. If a conflict is found the user is prompted to solve the conflict, if he/she is enabled to deal with entities at the top level, otherwise an action request is created in a queue for the top level master users. The same happens if the entity must be created and some new compatible entities are found.
The action request is a tuple $A = \langle ids, data, IID, message, history \rangle$ where $ids$ are the id of the involved entities in the top level EReg, $data$ is the new data received, $IID$ is the identifier of the instance sending such data, $message$ is the description of the issue, with contradictory, coincident and complementary attributes and relationships, and history is the current status of the request, the previous actions taken, involved users and so on. It is supposed that users in charge to deal with such request are allowed to query local documents where the data is found in order to check the authoritativeness of sources.

If the entity identified by $i_1$ is found in a document of another district, the process is similar: the first time a local entity is created and, when the data is sent to the top level EReg, the local EReg name and assigned id are appended to the existing entry. Partial identifiers management happens as above.

In the hypothesis that synchronization with the top level happens at fixed times, the basics of the process are the same, but the local updates generate a queue of synchronization actions. They are executed by a batch service, in the same way described above, with the difference that required user actions are written in a log similar to the action requests described above, and are not performed in real time.

\subsection{Query Mechanism}
\label{secquery}
The basic principle, leaving aside privacy issues for the moment, is as follows. When a user queries the system for an entity identified by some attribute set $a_1,...,a_k$, entities at local level are shown, with the documents where they are mentioned.
Then the top level EReg is queried; if the entity synchronization has already been performed as described above, all compatible candidates with their attributes, relationships and mentions are shown. The user can, accordingly, have an idea of all the entity in the whole system that are more or less compatible with the specified attribute set. On the other hand, if the architecture postpones entity synchronization at query time, the top level EReg queries all the local ERegs with the requested attribute set, performs the synchronization as described above and finally sends back results. As the process may require some time, the user does not get results in real time. In this scenario, data is accessed using a permission mechanisms based on the data described in section \ref{ssecdac}. When users access an \textbf{owned} document, they can see and possibly edit all the mentioned non public entities, their attributes and relationships derived from other owned documents and public entities with all attributes and relationships. If they have permissions to access documents at \textbf{office} or \textit{district} level, they will see also nonpublic entities in documents owned by the office or district, in the following ways, depending on the privacy - permission applicable entry:
\textit{i)} see entities, attributes, relationships and documents with their mentions and all other mentioned entities without restrictions, 
\textit{ii)} see entities, attributes, relationships and documents with their mentions and all other mentioned non public entities after anonymization,  \textit{iii)} see entities, attributes and relationships, but no mentions or documents; for instance they could learn the parents' names of a person, without being able to read the document containing the information,  \textit{iv)} entities without details, with the number of the mentioning documents, \textit{iv)} an error message, if their permissions are too low

More in details, suppose that users read a document $D_1$ which they own. All contained entities are shown without restrictions. If they choose one such entity, e.g. a specific person, they are able to see all its details and to navigate to all other documents where it is mentioned. Suppose they have just \textit{reader} rights on some such documents, containing mentions of a person $P_1$ never quoted in an owned document. If the \textit{Privacy - Permission} table entry for persons' privacy level and \textit{reader} ownership contains read anonymized as permission, they will read anonymized all $P_1$ details. More restrictive privacy level do not make sense for documents that the user is allowed to read.
If they have only \textit{generic} ownership on some such documents, and permission in the \textit{Privacy - Permission} table is without mentions, they will see only the quoted persons, without the text of the mentions; if it is count only, they will just learn how many persons are mentioned. Similar considerations apply if the users navigate the entity graph, that may be built using mentions. It has as nodes entities (with their attributes) and documents, with edges representing mentions. About documents at \textbf{other districts}, the behaviour is similar, but the permission check is performed by the top level EReg. It receives the user permission level with the query and uses the privacy - permission tables of each district owning the documents to know what the user can see of their entities. All the queries discussed so far, retrieve specific documents and entities. Other queries, even if based on entities, are statistics-oriented. For instance in \cite{bellandi2023} statistics of plaintiffs gender in divorce cases and of ages in job related cases are described. Such queries, in general, involve only counts of entities, without any details, and are not involved in privacy considerations.

\section{The Architecture}\label{secarch}
\begin{figure}
\centering
    \begin{subfigure}[b]{0.45\textwidth}
    \centering
    \includegraphics[width=0.99\textwidth]{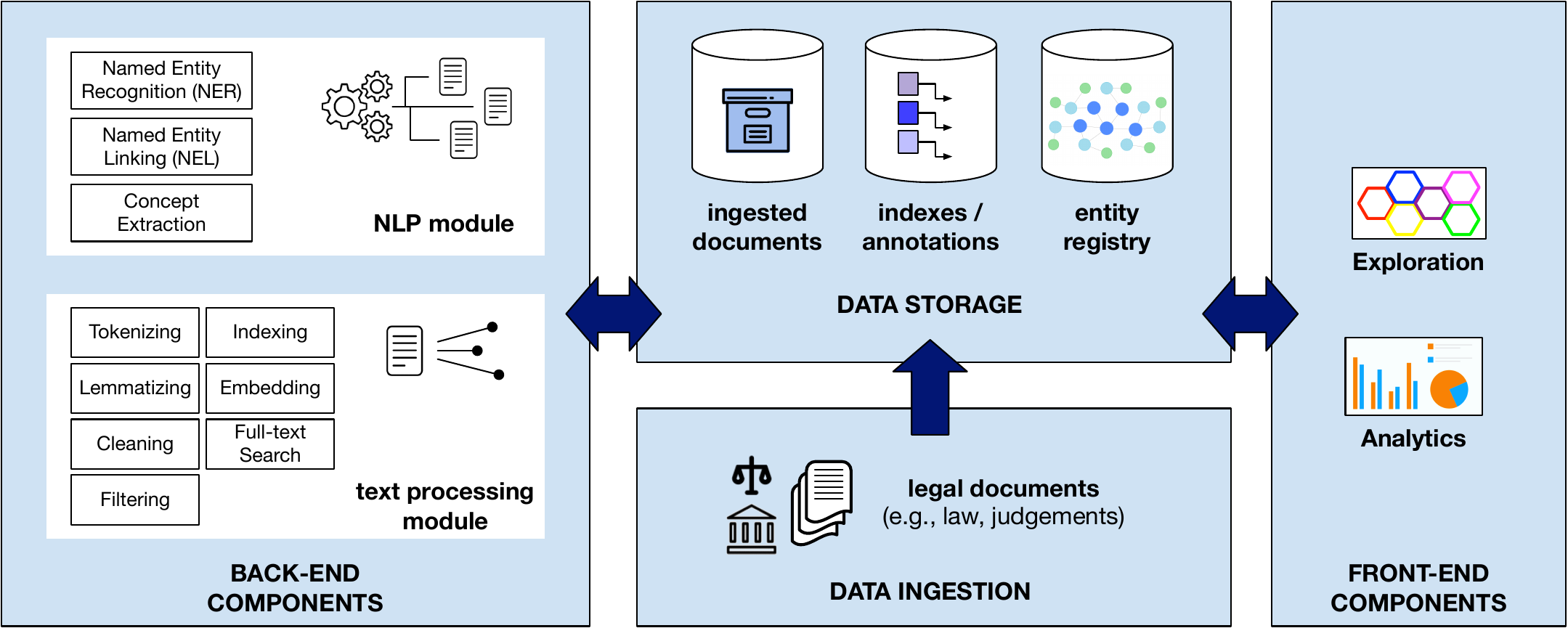}
    \caption{\label{fig1}}
    \end{subfigure}
\quad
    \begin{subfigure}[b]{0.32\textwidth}
    \centering
    \includegraphics[width=0.99\textwidth]{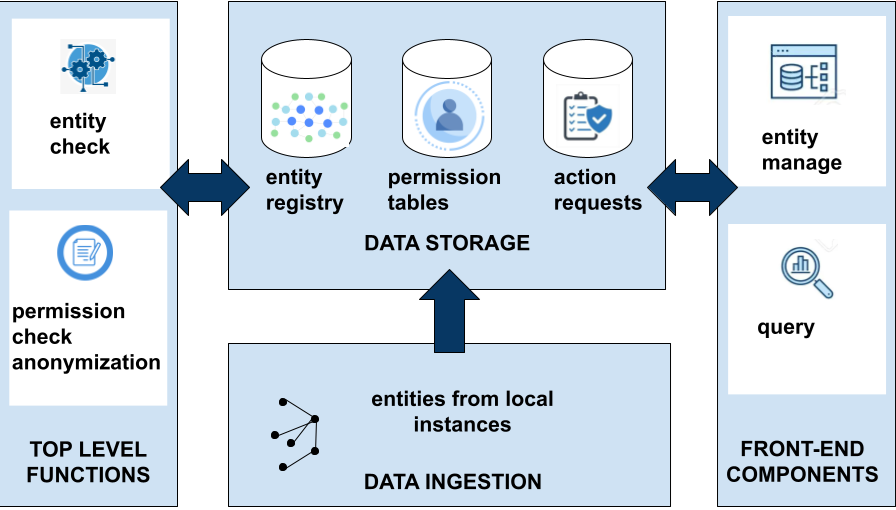}
    \caption{\label{fig2}}
    \end{subfigure}

\caption  { The district architecture description (\subref{fig1}) and  The top level architecture description (\subref{fig2})}
\end{figure}

The components of the architecture at each district are shown in Figure~\ref{fig1}. A {\em data ingestion} layer is defined to acquire the documents that needs to be managed. The documents are acquired progressively when they become available without system downtime. 
A {\em data storage layer} maintains the raw ingested documents and corresponding texts; the annotations as well as the index system for full text, metadata, and annotation search; and the graph database for the \textit{Entity Registry} (EReg) to store a unique entry for the entities extracted from documents. The storage layer exposes the EReg APIs to manage both the entity types (the EReg metamodel) and the entity instances as described in~\cite{dubai2023}. Document texts and metadata are stored in an ElasticSearch instance, while annotations in a SQL database as discussed in our previous work~\cite{smds}. As a graph database for EReg, we employed Neo4j.  In sight of the discussion of the hierarchical architecture, we recall that the EReg metamodel normally contains entries for \textit{Entity Types}, e.g. persons, cars, law articles, etc., their \textit{identifying attributes}, e.g. name, surname, personal code, etc. and \textit{identifiers}, that is sets of attributes that can uniquely identify an entity of the specified type. Identifiers are analogous to unique keys for SQL databases. Moreover, it contains attributes and relationships details. In \textit{back-end components}, we distinguish modules for: {\em text processing} to index and fetch data from the storage, to process the incoming data at ingestion time, and to create manipulated versions of the original documents through activities like segmentation, cleaning, and filtering; {\em NLP} to provide specific services according to the kind of mining operations that the system aims to support, like for example Named Entity Recognition (NER) and Linking (NEL), as well as concept extraction and statistics based on entities. 
An important subset of the NLP components is devoted to anonymization of documents, when they must be accessed by users not allowed to see all the mentioned entities (see section \ref{secquery}). Entity anonymization may be performed on the fly, and might also concern numeric data, if it is believed that they can identify involved people (see e.g. \cite{fabio2023} for a discussion in another context). All the NLP services must expose standard APIs for interaction with other platform components. In the end, the invoked NLP service passes back the output to the text processing module for storage in the annotation database and the entity registry. In {\em front-end components}, we extend our previous work in~\cite{smds} and we distinguish modules for {\em exploration} and {\em analytics}. These modules expose APIs to enforce the interaction of users with the back-end components. Exploration allows to move from one document to another according to similarity-based criteria. The idea is to provide a service for browsing the corpus according to their common entities and/or concepts extracted by the NLP module. Analytics allows to examine the corpus through summary/statistical views built over data, such as for example the distribution of an entity or concept in the corpus, the shortest path (through documents) between given concepts or entities, and the centrality of entities and concepts.

The top level architecture is described in fig. \ref{fig2}.
The \textit{data ingestion} layer is involved with entities only, as long as they are created or updated at local level. The \textit{data storage} contains the EReg with entities and metamodels; in this case, are included also tables listing, for each entity, all ids it has in the local instances where it is mentioned. Permission tables are stored to globally check user rights, when cross instance information is requested. Finally, requests to perform actions when inconsistencies are found are stored as described in section \ref{sectopl}.  The \textit{top level functions} layer implements functions described in section \ref{sectopl}, that is entity checks, to assess that no inconsistency arise when new pieces of information are ingested, and all checks and operations involved in permission checks as detailed in sections \ref{sectopl} and \ref{secquery}. Finally, \textit{front-end components} include tools to interactively manage entities to resolve inconsistencies and the query module. It is intended that the latter can be used interactively by users of the top level system, but its main goal is to serve requests coming from users of the local instances.

\section{Conclusion}
This paper presented a distributed architecture designed for efficiently managing legal documents and metadata. The system utilizes a decentralized network of nodes to analyze documents, providing advanced semantic management and improving scalability, fault tolerance, and performance. It was applied in the Italian Ministry of Justice to support the management of legal texts, enabling semantic exploration by extracting key insights from documents. The architecture also focuses on securely releasing data, ensuring privacy and protection of sensitive information through encryption, access controls, and anonymization. It addresses legal sector requirements, offering a scalable, privacy-conscious solution for managing and exploring legal documents, enhancing operational efficiency and data handling.

\section{Acknowledgments}
Research supported, in parts, by \emph{i)} Università degli Studi di Milano under the program ``Piano di Sostegno alla Ricerca''. \emph{ii)} project MUSA - Multilayered Urban Sustainability Action - project, funded by the European Union - NextGenerationEU, under the National Recovery and Resilience Plan (NRRP) Mission 4 Component 2 Investment Line 1.5: Strengthening of research structures and creation of R\&D ``innovation ecosystems'', set up of ``territorial leaders in R\&D'' (CUP  G43C22001370007, Code ECS00000037), \emph{iii)} project SERICS (PE00000014) under the NRRP MUR program funded by the EU - NGEU. Views and opinions expressed are however those of the authors only and do not necessarily reflect those of the European Union or the Italian MUR. Neither the European Union nor the Italian MUR can be held responsible for them.

\bibliography{references}

\begin{thebibliography}{18}
\expandafter\ifx\csname natexlab\endcsname\relax\def\natexlab#1{#1}\fi
\providecommand{\url}[1]{\texttt{#1}}
\providecommand{\href}[2]{#2}
\providecommand{\path}[1]{#1}
\providecommand{\DOIprefix}{doi:}
\providecommand{\ArXivprefix}{arXiv:}
\providecommand{\URLprefix}{URL: }
\providecommand{\Pubmedprefix}{pmid:}
\providecommand{\doi}[1]{\href{http://dx.doi.org/#1}{\path{#1}}}
\providecommand{\Pubmed}[1]{\href{pmid:#1}{\path{#1}}}
\providecommand{\bibinfo}[2]{#2}
\ifx\xfnm\relax \def\xfnm[#1]{\unskip,\space#1}\fi
\bibitem[{Pauzi and Capiluppi(2023)}]{PAUZI2023111616}
\bibinfo{author}{Z.~Pauzi}, \bibinfo{author}{A.~Capiluppi},
\newblock \bibinfo{title}{Applications of natural language processing in software traceability: A systematic mapping study},
\newblock \bibinfo{journal}{Journal of Systems and Software} \bibinfo{volume}{198} (\bibinfo{year}{2023}) \bibinfo{pages}{111616}. \URLprefix \url{https://www.sciencedirect.com/science/article/pii/S0164121223000110}. \DOIprefix\doi{https://doi.org/10.1016/j.jss.2023.111616}.
\bibitem[{Amato et~al.(2008)Amato, Mazzeo, Penta, and Picariello}]{amato2008}
\bibinfo{author}{F.~Amato}, \bibinfo{author}{A.~Mazzeo}, \bibinfo{author}{A.~Penta}, \bibinfo{author}{A.~Picariello},
\newblock \bibinfo{title}{Using nlp and ontologies for notary document management systems},
\newblock in: \bibinfo{booktitle}{Proceedings of the 2008 19th International Conference on Database and Expert Systems Application}, DEXA '08, \bibinfo{publisher}{IEEE Computer Society}, \bibinfo{address}{USA}, \bibinfo{year}{2008}, p. \bibinfo{pages}{67–71}. \URLprefix \url{https://doi.org/10.1109/DEXA.2008.86}. \DOIprefix\doi{10.1109/DEXA.2008.86}.
\bibitem[{Humphreys et~al.(2021)Humphreys, Boella, van~der Torre, Robaldo, Caro, Ghanavati, and Muthuri}]{humphreys2020}
\bibinfo{author}{L.~Humphreys}, \bibinfo{author}{G.~Boella}, \bibinfo{author}{L.~van~der Torre}, \bibinfo{author}{L.~Robaldo}, \bibinfo{author}{L.~D. Caro}, \bibinfo{author}{S.~Ghanavati}, \bibinfo{author}{R.~Muthuri},
\newblock \bibinfo{title}{Populating legal ontologies using semantic role labeling},
\newblock \bibinfo{journal}{Artificial Intelligence and Law} \bibinfo{volume}{29} (\bibinfo{year}{2021}) \bibinfo{pages}{171--211}.
\bibitem[{Bellandi et~al.(2024)Bellandi, Bernasconi, Lodi, Palmonari, Pozzi, Ripamonti, and Siccardi}]{BELLANDI2024105904}
\bibinfo{author}{V.~Bellandi}, \bibinfo{author}{C.~Bernasconi}, \bibinfo{author}{F.~Lodi}, \bibinfo{author}{M.~Palmonari}, \bibinfo{author}{R.~Pozzi}, \bibinfo{author}{M.~Ripamonti}, \bibinfo{author}{S.~Siccardi},
\newblock \bibinfo{title}{An entity-centric approach to manage court judgments based on natural language processing},
\newblock \bibinfo{journal}{Computer Law \& Security Review} \bibinfo{volume}{52} (\bibinfo{year}{2024}) \bibinfo{pages}{105904}. \URLprefix \url{https://www.sciencedirect.com/science/article/pii/S0267364923001140}. \DOIprefix\doi{https://doi.org/10.1016/j.clsr.2023.105904}.
\bibitem[{Qin et~al.(2024)Qin, Chen, and Mou}]{QIN2024105930}
\bibinfo{author}{H.~Qin}, \bibinfo{author}{L.~Chen}, \bibinfo{author}{L.~Mou},
\newblock \bibinfo{title}{The development of china's electronic case file regulations and its future implications},
\newblock \bibinfo{journal}{Computer Law \& Security Review} \bibinfo{volume}{52} (\bibinfo{year}{2024}) \bibinfo{pages}{105930}. \URLprefix \url{https://www.sciencedirect.com/science/article/pii/S0267364923001401}. \DOIprefix\doi{https://doi.org/10.1016/j.clsr.2023.105930}.
\bibitem[{Ridwandono et~al.(2023)Ridwandono, Afandi, Wahyuni, Simaremare, and Sinaga}]{Ridwandono_Afandi_Wahyuni_Simaremare_Sinaga_2023}
\bibinfo{author}{D.~Ridwandono}, \bibinfo{author}{M.~I. Afandi}, \bibinfo{author}{E.~D. Wahyuni}, \bibinfo{author}{E.~Simaremare}, \bibinfo{author}{I.~Sinaga},
\newblock \bibinfo{title}{Legal documents repository systems},
\newblock \bibinfo{journal}{Nusantara Science and Technology Proceedings} \bibinfo{volume}{2023} (\bibinfo{year}{2023}) \bibinfo{pages}{477--481}. \URLprefix \url{https://nstproceeding.com/index.php/nuscientech/article/view/983}. \DOIprefix\doi{10.11594/nstp.2023.3377}.
\bibitem[{Dongarra et~al.(2019)Dongarra, Tourancheau, Balouek-Thomert, Renart, Zamani, Simonet, and Parashar}]{10.1177/1094342019877383}
\bibinfo{author}{J.~Dongarra}, \bibinfo{author}{B.~Tourancheau}, \bibinfo{author}{D.~Balouek-Thomert}, \bibinfo{author}{E.~G. Renart}, \bibinfo{author}{A.~R. Zamani}, \bibinfo{author}{A.~Simonet}, \bibinfo{author}{M.~Parashar},
\newblock \bibinfo{title}{Towards a computing continuum: Enabling edge-to-cloud integration for data-driven workflows},
\newblock \bibinfo{journal}{Int. J. High Perform. Comput. Appl.} \bibinfo{volume}{33} (\bibinfo{year}{2019}) \bibinfo{pages}{1159–1174}. \URLprefix \url{https://doi.org/10.1177/1094342019877383}. \DOIprefix\doi{10.1177/1094342019877383}.
\bibitem[{Dos~Anjos et~al.(2020)Dos~Anjos, Matteussi, De~Souza, Grabher, Borges, Barbosa, González, Leithardt, and Geyer}]{9193894}
\bibinfo{author}{J.~C.~S. Dos~Anjos}, \bibinfo{author}{K.~J. Matteussi}, \bibinfo{author}{P.~R.~R. De~Souza}, \bibinfo{author}{G.~J.~A. Grabher}, \bibinfo{author}{G.~A. Borges}, \bibinfo{author}{J.~L.~V. Barbosa}, \bibinfo{author}{G.~V. González}, \bibinfo{author}{V.~R.~Q. Leithardt}, \bibinfo{author}{C.~F.~R. Geyer},
\newblock \bibinfo{title}{Data processing model to perform big data analytics in hybrid infrastructures},
\newblock \bibinfo{journal}{IEEE Access} \bibinfo{volume}{8} (\bibinfo{year}{2020}) \bibinfo{pages}{170281--170294}. \DOIprefix\doi{10.1109/ACCESS.2020.3023344}.
\bibitem[{Ardagna et~al.(2015)Ardagna, Asal, Damiani, and Vu}]{AADV.CSUR2015}
\bibinfo{author}{C.~Ardagna}, \bibinfo{author}{R.~Asal}, \bibinfo{author}{E.~Damiani}, \bibinfo{author}{Q.~Vu},
\newblock \bibinfo{title}{{From Security to Assurance in the Cloud: A Survey}},
\newblock \bibinfo{journal}{ACM CSUR} \bibinfo{volume}{48} (\bibinfo{year}{2015}).
\bibitem[{Ardagna and Bena(2023)}]{AB.SSE2023}
\bibinfo{author}{C.~A. Ardagna}, \bibinfo{author}{N.~Bena},
\newblock \bibinfo{title}{Non-functional certification of modern distributed systems: A research manifesto},
\newblock in: \bibinfo{booktitle}{Proc. of IEEE SSE 2023}, \bibinfo{address}{Chicago, IL, USA}, \bibinfo{year}{2023}.
\bibitem[{Zhong et~al.(2020)Zhong, Xiao, Tu, Zhang, Liu, and Sun}]{zhong2020does}
\bibinfo{author}{H.~Zhong}, \bibinfo{author}{C.~Xiao}, \bibinfo{author}{C.~Tu}, \bibinfo{author}{T.~Zhang}, \bibinfo{author}{Z.~Liu}, \bibinfo{author}{M.~Sun}, \bibinfo{title}{How does {NLP} benefit legal system: A summary of legal artificial intelligence}, \bibinfo{year}{2020}.
\bibitem[{Chalkidis et~al.(2020)Chalkidis, Fergadiotis, Malakasiotis, Aletras, and Androutsopoulos}]{chalkidis-etal-2020-legal}
\bibinfo{author}{I.~Chalkidis}, \bibinfo{author}{M.~Fergadiotis}, \bibinfo{author}{P.~Malakasiotis}, \bibinfo{author}{N.~Aletras}, \bibinfo{author}{I.~Androutsopoulos},
\newblock \bibinfo{title}{{LEGAL}-{BERT}: The muppets straight out of law school},
\newblock in: \bibinfo{booktitle}{Findings of the Association for Computational Linguistics: EMNLP 2020}, \bibinfo{publisher}{Association for Computational Linguistics}, \bibinfo{address}{Online}, \bibinfo{year}{2020}.
\bibitem[{Castano et~al.(2022)Castano, Falduti, Ferrara, and Montanelli}]{CASTANO2022101842}
\bibinfo{author}{S.~Castano}, \bibinfo{author}{M.~Falduti}, \bibinfo{author}{A.~Ferrara}, \bibinfo{author}{S.~Montanelli},
\newblock \bibinfo{title}{A knowledge-centered framework for exploration and retrieval of legal documents},
\newblock \bibinfo{journal}{Information Systems} \bibinfo{volume}{106} (\bibinfo{year}{2022}).
\bibitem[{Rabelo et~al.(2022)Rabelo, Goebel, and Kim}]{cooliedescropt}
\bibinfo{author}{J.~Rabelo}, \bibinfo{author}{R.~Goebel}, \bibinfo{author}{M.-Y. e.~a. Kim},
\newblock \bibinfo{title}{Overview and discussion of the competition on legal information extraction/entailment (coliee) 2021},
\newblock \bibinfo{journal}{The Review of Socionetwork Strategies} \bibinfo{volume}{16} (\bibinfo{year}{2022}).
\bibitem[{Bellandi et~al.(2023)Bellandi, Maghool, and Siccardi}]{bellandi2023}
\bibinfo{author}{V.~Bellandi}, \bibinfo{author}{S.~Maghool}, \bibinfo{author}{S.~Siccardi},
\newblock \bibinfo{title}{An nlp-based statistical reporting methodology applied to court decisions},
\newblock in: \bibinfo{booktitle}{2023 49th Euromicro Conference on Software Engineering and Advanced Applications (SEAA)}, \bibinfo{publisher}{IEEE Computer Society}, \bibinfo{address}{Los Alamitos, CA, USA}, \bibinfo{year}{2023}, pp. \bibinfo{pages}{108--111}. \URLprefix \url{https://doi.ieeecomputersociety.org/10.1109/SEAA60479.2023.00025}. \DOIprefix\doi{10.1109/SEAA60479.2023.00025}.
\bibitem[{Bellandi and Siccardi(2023)}]{dubai2023}
\bibinfo{author}{V.~Bellandi}, \bibinfo{author}{S.~Siccardi},
\newblock \bibinfo{title}{An entity registry: A model for a repository of entities found in a document set},
\newblock in: \bibinfo{booktitle}{NIAI, MoWiN, AIAP, SIGML, CNSA, ICCIoT - 2023}, \bibinfo{organization}{AIRCC Publishing Corporation}, \bibinfo{year}{2023}, p. \bibinfo{pages}{1–12}.
\bibitem[{Batini et~al.(2021)Batini, Bellandi, Ceravolo, Moiraghi, Palmonari, and Siccardi}]{smds}
\bibinfo{author}{C.~Batini}, \bibinfo{author}{V.~Bellandi}, \bibinfo{author}{P.~Ceravolo}, \bibinfo{author}{F.~Moiraghi}, \bibinfo{author}{M.~Palmonari}, \bibinfo{author}{S.~Siccardi},
\newblock \bibinfo{title}{Semantic data integration for investigations: Lessons learned and open challenges},
\newblock in: \bibinfo{booktitle}{2021 IEEE International Conference on Smart Data Services (SMDS)}, \bibinfo{year}{2021}.
\bibitem[{Giampaolo et~al.(2023)Giampaolo, Izzo, Siccardi, Polimeno, Bellandi, and Piccialli}]{fabio2023}
\bibinfo{author}{F.~Giampaolo}, \bibinfo{author}{S.~Izzo}, \bibinfo{author}{S.~Siccardi}, \bibinfo{author}{A.~Polimeno}, \bibinfo{author}{V.~Bellandi}, \bibinfo{author}{F.~Piccialli},
\newblock \bibinfo{title}{Real-time anonymization of sensitive personal data using a service-based architecture},
\newblock in: \bibinfo{booktitle}{2023 IEEE International Conference on Web Services (ICWS)}, \bibinfo{year}{2023}, pp. \bibinfo{pages}{701--703}.

\end{thebibliography}
\end{document}